\documentclass{epl}

\title{Superfluidity in a gas of strongly-interacting bosons}
\author{G. M. Kavoulakis, Y. Yu, M. \"Ogren, and S. M. Reimann}
\institute{Mathematical Physics, LTH, Lund University, P.O. Box
118, S-22100 Lund, Sweden}
\begin{document}

\maketitle

\begin{abstract}
We consider small systems of bosonic atoms rotating in a
toroidal trap. Using the method of exact numerical
diagonalization of the many-body Hamiltonian, we examine the
transition from the Bose-Einstein condensed state to the
Tonks-Girardeau state. The system supports persistent currents
in a wide range between the two limits, even in the absence of
Bose-Einstein condensation.
\end{abstract}

\section{Introduction} The advances in experimental techniques
in the field of cold atoms have made it possible to access
experimentally strongly correlated states that had only been
part of the mathematical apparatus of quantum theory up to now.
A famous recent example is the Tonks-Girardeau limit of bosonic
atoms \cite{TGIR} that has been realized in the two
experimental studies of Refs.\,\cite{Science}. Confined in
elongated, quasi one-dimensional traps -- as it nowadays can be
realized in many laboratories around the world -- at low
densities or strong interparticle interactions the boson gas
exhibits some properties which resemble those of a fermionic
system.

Here, we consider interacting bosonic atoms which rotate in a
toroidal trap. We examine the transition from the so-called
Gross-Pitaevskii limit of weak interactions, where the gas
forms a Bose-Einstein condensate, to the Tonks-Girardeau limit.
In this transition, while Bose-Einstein condensation
disappears, superfluidity persists in a much wider region.
``Superfluidity", throughout this paper, means the
metastability of currents, or in other words, the existence of
persistent currents \cite{Sonin,Astr,Brand,decay}, which is
defined more precisely below.

The whole question of superfluidity is based upon two crucial
issues: firstly, that there is an energy barrier between the
current-carrying state and the current-free state. Secondly
that no single-particle operator $\Delta V$ connects the two
states, at least to some order. In this case the decay rate of
the current due to some weak perturbation that dissipates
angular momentum and energy, if necessary, is highly
suppressed. In the present problem the criterion we use is that
no single-particle operator connects the two states at least to
{\it first-order} in this perturbation. As we will see, the two
states are connected via $\Delta V$ to lower and lower order,
as the system goes from the Gross-Pitaevskii to the
Tonks-Girardeau regimes. As a result, the characteristic
timescale of the decay decreases in the transition between
these two limits. Still, a transition that is not allowed even
to first-order, will take a very long timescale (compared to
the lifetime of these gases) for sufficiently weak
perturbations. This approach is equivalent to examining the
dynamic structure factor, as in Refs.\,\cite{Astr,Brand}.

It is known that Bose-Einstein condensation is neither a
necessary, nor a sufficient condition for superfluidity
\cite{Leggett}. The problem investigated here provides an
example where: (i) for Bose-Einstein condensation to imply
superfluidity, interactions are necessary, and (ii)
superfluidity exists in a much wider region between the two
limits, even in the absence of Bose-Einstein condensation.

Assuming the usual contact interactions between the atoms, the
many-body Hamiltonian is
\begin{eqnarray}
  {\hat H} = - \frac {\hbar^2} {2 M} \int \psi^{\dagger}({\bf r})
  \nabla^2 \psi({\bf r}) \, d {\bf r} +
   \frac {U_0} {2} \int \psi^{\dagger}({\bf r})
  \psi^{\dagger}({\bf r}) \psi({\bf r}) \psi({\bf r}) \, d {\bf r}.
\label{Ham1}
\end{eqnarray}
Here $M$ is the atom mass and $U_0 = 4 \pi \hbar^2 a /M$, where
$a$ is the scattering length for elastic atom-atom collisions.
Since we assume that the atoms are confined in a toroidal trap,
we express the bosonic annihilation operator $\psi$ in terms of
the annihilation operators $c_m$ of an atom with angular
momentum $m \hbar$, $\psi({\bf r}) = \sum_m c_m e^{i m \theta}
/ {\sqrt V}$. Here $V = 2 \pi R S$ is the volume of the torus
with $R$ its radius and $S$ its cross section, with $\sqrt S
\ll R$. Measuring the energy in units of $E_0 = \hbar^2 / (2 M
R^2)$ we thus find \cite{Lieb1,Lieb2}
\begin{equation}
  {\hat H} = \sum_l l^2 c_l^{\dagger} c_l
      + \frac {g} {2} \sum_{k,l,m,n} c_k^{\dagger} c_l^{\dagger}
      c_m c_n \, \delta_{k+l,m+n},
\label{Ham4}
\end{equation}
where the dimensionless constant $g$ equals $4 a R/S$.

\section{Transition between the Gross-Pitaevskii and the
Tonks-Girardeau regimes} In the Gross-Pitaevskii limit one has
a Bose-Einstein condensate. In this case, the typical kinetic
energy per atom is $\hbar^2/(2MR^2)$, and the typical
interaction energy per atom is $n U_0$. Here $n = N/V$ is the
atom density and $N$ is the atom number. From these two energy
scales, we see that in this limit $g$ should not exceed $g_{\rm
GP} \sim 1/N$. In the opposite Tonks-Girardeau extreme, the
typical kinetic energy per atom is on the order of $\hbar^2
k_F^2 / (2 M)$ (just like the Fermi energy in a Fermi gas),
with $k_F = \sigma \pi$ and $\sigma = N /(2 \pi R)$ the density
per unit length. The interaction energy is still on the order
of $n U_0$, and thus the characteristic value of $\sigma$ is of
order $a/S$. Therefore, the coefficient $a R/S$ is of order $N$
in the Tonks-Girardeau limit, $g \ge g_{\rm TG} \sim N$. The
parameter $\gamma$ in the Lieb-Liniger model \cite{Lieb1} is
$(\pi/2) g /N$. Then, for $g \sim N$, $\gamma$ is indeed of
order unity. The above estimates are in agreement with the two
higher curves of Fig.\,2, as explained below.

\section{Gross-Pitaevskii limit and persistent currents}
Although the results of the numerical diagonalization that we
present below are much more general, it is instructive to start
with a mean-field variational calculation that is valid in the
Gross-Pitaevskii extreme. The existence of persistent currents
can be seen easily in the following way
\cite{Rokhsar,Leggett,Putterman}. Let us start with the usual
basis states of angular momentum $m \hbar$, $\phi_m (\theta)=
{\sqrt N} e^{i m \theta}/{\sqrt{2 \pi R}}$ (normalized to $N$),
and consider a linear superposition of $\phi_0$ and $\phi_1$,
$\psi_{\rm var}(\theta) = c_0 \, \phi_0 + c_1 \, \phi_1 =
\sqrt{1-l} \,\phi_0 + e^{i \phi} {\sqrt l} \, \phi_1$. Here
$\phi$ is an arbitrary phase, and $l$ is a parameter, being
equal to the expectation value of the angular momentum per
particle, $l=L/N$. For $l=1$, $\psi_{\rm var}$ describes a
vortex located at its center, while for $l=0$ the vortex is at
an infinite distance away from the torus.

Because of the extra kinetic energy, the expectation value of
the energy in the above state is higher for $l=1$ than for
$l=0$. The first decisive question is whether there exists an
energy barrier for some intermediate value of $l$. Such a
barrier is expected to be present for sufficiently strong
repulsive interatomic interactions for the following reason: As
$l$ decreases, at some intermediate value between zero and
unity, the vortex passes through the torus, which implies that
a node forms in the atom density. This, however, costs
interaction energy, since for $l=0$ and $l=1$ the density is
homogeneous, which is the configuration with the lowest
interaction energy. Therefore, for sufficiently strong
interactions, such a barrier is expected to develop.

Quantitatively, the expectation value of the Hamiltonian in the
state $\psi_{\rm var}$ above is the Gross-Pitaevskii energy
$E_{\rm GP}$, which is (in units of $E_0$)
\begin{eqnarray}
   {E_{\rm GP}}/N - {\delta}/2 = (1+\delta) l - \delta l^2 +
   {\cal O}(g),
\label{Ham8}
\end{eqnarray}
with $\delta = (N-1)g$ \cite{slope}. According to
Eq.\,(\ref{Ham8}), $E_{\rm GP}$ develops a local maximum at
$l=(1+\delta)/(2 \delta)$. In order for this to appear at a
value of $l$ smaller than unity, $\delta > 1$, or $g>1/(N-1)$.
Therefore, this is the critical value of $\delta$ (and of $g$)
above which an energy barrier develops in the dispersion
relation $E_{\rm GP}=E_{\rm GP}(l)$. Equation (\ref{Ham8}) also
coincides with the excitation spectrum calculated by Lieb
\cite{Lieb2} in the limit $N \to \infty$ and $R \to \infty$,
with $N/(2 \pi R)=\sigma$, finite, and for small $\gamma$,
i.e., $E_{\rm GP}/N - 4 \pi \gamma= 8 \pi \gamma (l - l^2)$ in
the units of energy of Refs.\,\cite{Lieb1,Lieb2}, $\hbar^2
\sigma^2/2M$.

In order for our argument for the persistent currents to be
complete, we need to consider the possibility of connecting the
many-body states $\Psi_0$ with $L=0$ and $\Psi_N$ with $L=N$
via some single-particle operator $\Delta V$ of the form
$\Delta V = V_0 \delta (\theta)$. Only one Fourier component of
the delta function connects the two states and provides the
difference in the angular momentum between them, thus
\begin{eqnarray}
   \Delta V = V_0 \sum_{j=1}^N e^{i N \theta_j} + c.c.
\end{eqnarray}
The above operator is precisely the one that enters in the
calculation of the dynamic structure factor $S(k,\omega)$ for
$k = 2 k_F$ (or $L=N$ in our model). In Ref.\,\cite{Brand} it
has been argued that the vanishing of $S(k=2k_F, \omega=0)$
implies that the decay rate of persistent currents is
suppressed, in agreement with the present study, as we discuss
below.

The state $\Psi_{L=N}$ results from $\Psi_{L=0}$ as $\Psi_{L=N}
= \left[ \prod_{j=1}^N e^{i \theta_j} \right] \Psi_{L=0}$,
(i.e., by exciting the center of mass) \cite{Bloch}, and thus
\begin{eqnarray}
    \langle \Psi_{L = 0} | \Delta V | \Psi_{L = N} \rangle =
  N V_0 \int
   |\Psi_{L=0}|^2 e^{i[-N \theta_1 + \theta_1 + \dots + \theta_N]}
  \, d \theta_1 \dots d\theta_N.
\end{eqnarray}
In the Gross-Pitaevskii limit $|\Psi_{L=0}|$ is independent of
$\theta_j$, and the matrix element vanishes; only high-order
perturbation theory in $\Delta V$ then connects the states
$|\Psi_{L=0} \rangle$ and $|\Psi_{L=N} \rangle$.

\section{Exact results from numerical diagonalization of the
Hamiltonian} To go beyond the mean-field approximation (which
breaks down far away from the Gross-Pitaevskii limit of weak
interactions), we have performed numerical diagonalization of
the full many-body Hamiltonian $\hat H$. We thus evaluate
numerically the eigenstates of $\hat H$, which are also
eigenstates of the momentum operator $\hat L$, and of the
number operator $\hat N$, for various values of $g$.

Although ``exact" (apart from the naturally necessary
truncation of the Hilbert space), this method is quite
restrictive in the values of $N$, $L$, and $g$ that can be
considered. More specifically, the number of the basis states
that need to be included increases as one approaches the
Tonks-Girardeau regime. In the Gross-Pitaevskii limit for a
non-rotating cloud the dominant contribution to the many-body
state comes from the $m=0$ single-particle state. In the
opposite extreme of the Tonks-Girardeau limit, even for a
non-rotating cloud, the dominant contribution comes from the
single-particle states with $|m| < m_{\rm max} \sim N/2$, since
in this limit the many-body wavefunction has to build
correlations of characteristic length $R/N$. Furthermore, for a
given interaction strength $U_0$ and a given density $n$,
$\hbar^2 m_{\rm max}^2/(2M R^2) \sim n U_0$, or $m_{\rm max}^2
\sim Ng$. Thus, more generally, $m_{\rm max} \sim {\rm max}
(N/2, \sqrt{Ng})$.

\begin{figure}
\onefigure[width=6.5cm,height=3.6cm]{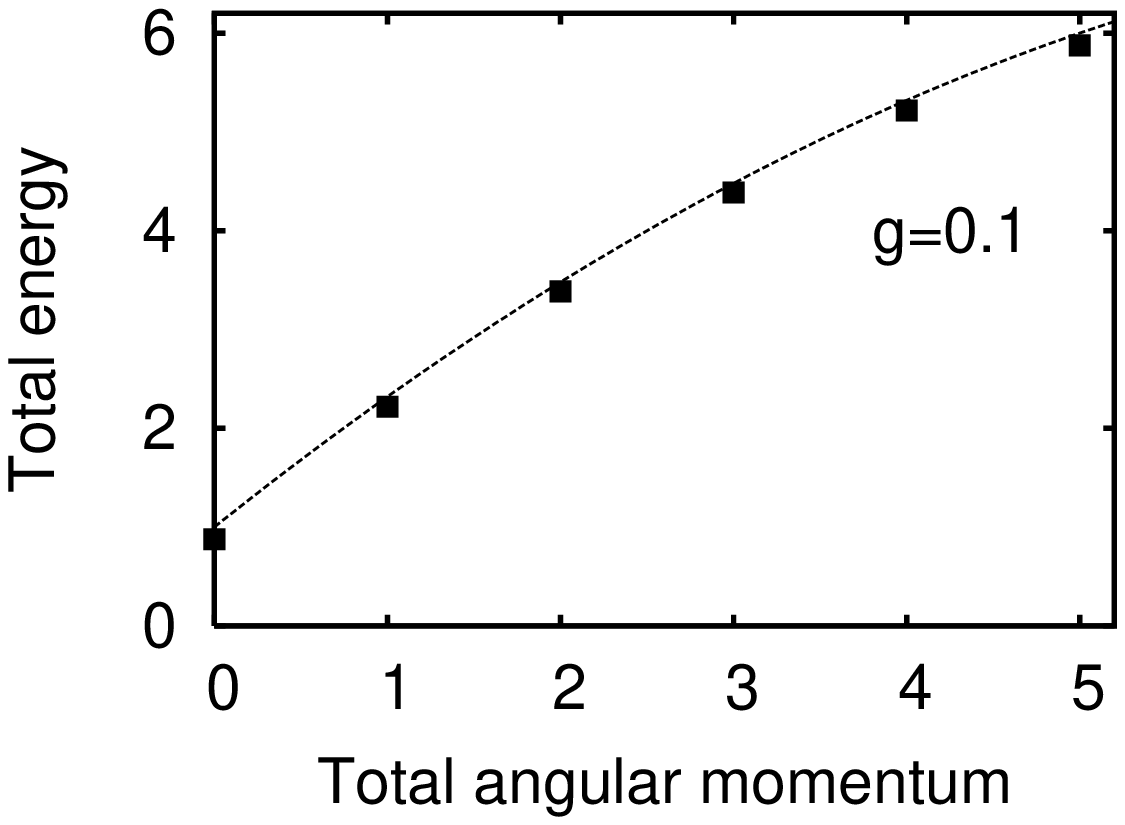}
\onefigure[width=6.5cm,height=3.6cm]{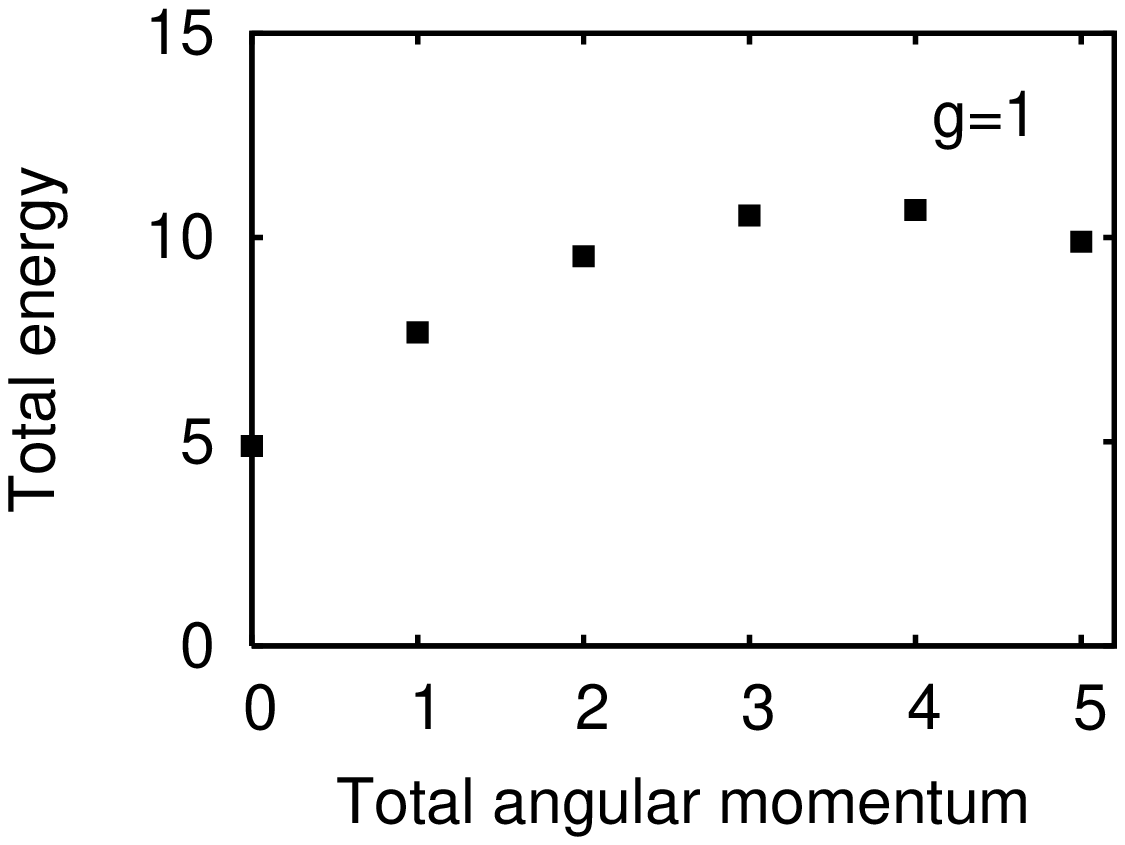}
\onefigure[width=6.5cm,height=3.6cm]{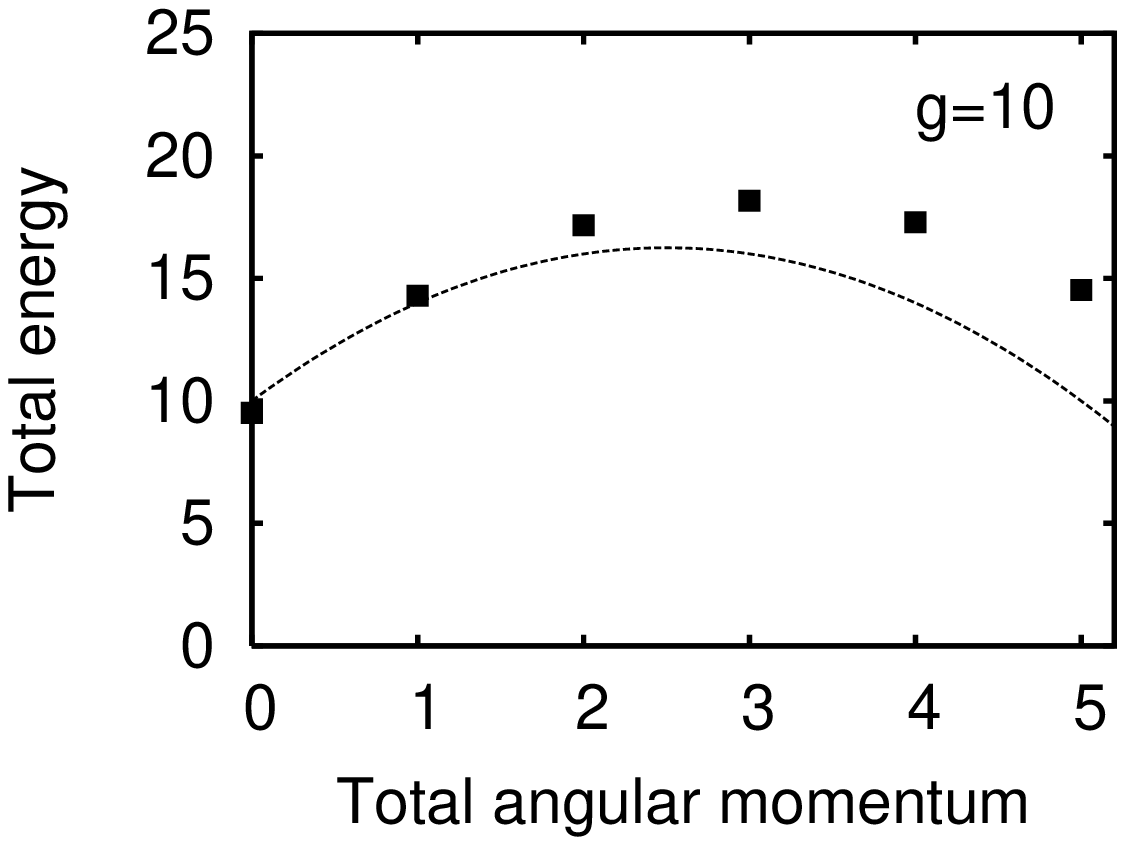} \caption{The
points show the lowest eigenenergies of the Hamiltonian ${\cal
E}={\cal E}(L)$ for $L=0,\dots,5$ and $N=5$. Also $g=0.1$
(top), $g=1$ (middle), and $g=10$ (bottom). The curves show the
energy as given by the mean-field approximation $E_{\rm GP}$
[Eq.\,(\ref{Ham8})] in the top graph and $E_{\rm TG}$
[Eq.\,(\ref{etg})] in the bottom graph.}
\label{FIG1}
\end{figure}

In Fig.\,1 we consider $N=5$ atoms, and $L=0,\dots,5$, with
$m_{\rm max} = 20$, which gives a change of the lowest
eigenenergy that is less than $1 \%$ between $m_{\rm max}=19$
and $m_{\rm max}=20$, for the highest value of $g=10$ that we
used. For $N=5$, $g_{\rm GP} \sim 1/N = 0.2$, while $g_{\rm TG}
\sim N =5$. Figure 1 shows the lowest eigenenergy ${\cal E}$ of
the Hamiltonian as function of the total angular momentum $L$,
for $g=0.1$, $g=1$, and $g=10$. As argued above, for $g=0.1$
the system is essentially in the Gross-Pitaevskii limit.
Furthermore, there is no maximum in ${\cal E}={\cal E}(L)$,
since the interaction is not strong enough (the mean-field
approximation predicts that the critical value of $g$ is 0.25
for 5 atoms). The dashed curve in the upper graph of Fig.\,1
shows the result of the mean-field approximation. Clearly, it
is possible to improve the mean-field energies by including
states with higher $|m|$ in $\psi_{\rm var}$. For $g=1$, there
is a maximum in ${\cal E}={\cal E}(L)$ at $L=4$. Finally, for
$g=10$, the system is closer to the Tonks-Girardeau limit, and
there is still a maximum in ${\cal E}={\cal E}(L)$. The dashed
curve in the lower graph of Fig.\,1 is the energy predicted in
the Tonks-Girardeau limit \cite{Lieb2},
\begin{eqnarray}
E_{\rm TG} = N^3/12 - L^2 + N L,
\label{etg}
\end{eqnarray}
plus corrections of order $N$. Such corrections are seen
clearly in this graph. For $L=0$ the leading-order term in the
energy is $N^3 /12$, which is the total energy of a Fermi gas
in one dimension, $N E_F/3$ (with $E_F$ the Fermi energy), for
a value of the Fermi momentum $k_F = \sigma \pi = N/(2R)$. The
expression for $E_{\rm TG}$ implies that the energy barrier in
${\cal E = \cal E}(L)$ [i.e., the difference between the
maximum in ${\cal E}(L)$ and ${\cal E}(L=N)$] scales
quadratically with $N$ in the Tonks-Girardeau limit, in
agreement with our numerical results. On the other hand, in the
Gross-Pitaevskii limit, the maximum of the energy barrier
scales linearly with $N$. This is the reason for the emergence
of this barrier as one goes from the one limit to the other.

\section{An exact result} The eigenenergies calculated above
satisfy an exact formula, valid for any value of $g$, which is
in agreement with the study of Felix Bloch \cite{Bloch}. More
specifically, ${\cal E}(L') - {\cal E}(L)= L'-L$, which also
implies that ${\cal E}(L=N)-{\cal E}(L=0)= N$, or more
generally ${\cal E}(L=qN)- {\cal E}(L=0)=q^2N$, $q =
0,1,2,\dots$ Geometrically ${\cal E}(L') - {\cal E}(L)= L'-L$
implies that all the lines connecting the points with $L$ and
$L'= N-L$ are parallel.

To see this result, we recall that any state $|L\rangle$ is
related to the state $|L'=N-L \rangle$ via some collective
excitation of the system. The state $|L\rangle$ is a linear
superposition of Fock states of the form $|(-m)^{N_{-m}},
\dots, 0^{N_0}, \dots, m^{N_m} \rangle$, with $\sum_m N_m = N$,
and $\sum_m m N_m = L$. Then, the state that consists of the
Fock states $|(-m+1)^{N_{m}}, \dots, (-1)^{N_2}, 0^{N_{1}},
1^{N_0}, \dots, m^{N_{-m+1}}, (m+1)^{N_{-m}} \rangle$ {\it with
the same amplitudes} as $|L \rangle$ is $|L'=N-L \rangle$,
since it has $N$ atoms, $L' = \sum_m (-m+1) N_m = N - L$, and a
total energy which is higher than that of $|L \rangle$ by
$\sum_m [(-m+1)^2-m^2] N_{m}= L'-L$ purely because of the
kinetic energy; the interaction energy is the same in
$|L\rangle$ and $|L'\rangle$, since all the matrix elements of
the interaction are the same, independent of $m$. Similar
arguments apply in all the branches with $pN \le L, L' \le
(p+1)N$, with $p=1,2,3,\dots$

\begin{figure}
\onefigure[width=7.5cm,height=4.7cm]{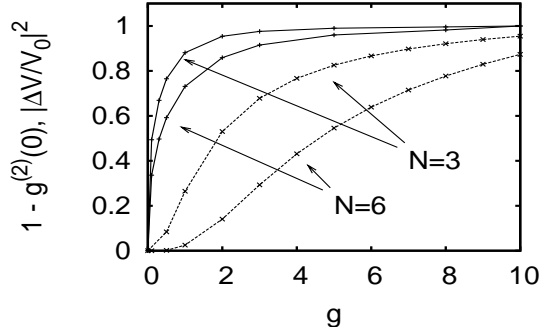} \caption{Two
higher curves: the function $1 - g^{(2)}(0)$, for $N=3$ and
$N=6$ atoms. Two lower curves: the matrix element $|\langle
\Psi_{L = 0} | \Delta V | \Psi_{L = N} \rangle/V_0|^2$ for
$N=3$ and $N=6$ atoms.}
\label{FIG2}
\end{figure}

\section{Persistent currents close to the Tonks-Girardeau
limit} In this last part of our study we examine the lifetime
of the current-carrying state with $L=N$ close to the
Tonks-Girardeau limit. Due to the Bose-Fermi mapping
\cite{TGIR}, for $\gamma \to \infty$ the matrix element
$\langle \Psi_{L = 0} | \Delta V | \Psi_{L = N} \rangle$ is
identical to the one for non-interacting fermions and equal to
$V_0$. The two lower curves in Fig.\,2 show $|\langle \Psi_{L =
0} | \Delta V | \Psi_{L = N} \rangle/V_0|^2$, for $N=3$ and
$N=6$, which indeed tend to unity for large $g$. Therefore, for
$\gamma \to \infty$ there is a single-particle operator that
connects the two states and destroys the persistent current for
$L=N$. In addition, the difference in the energy associated
with the transition between the states $\Psi_{L=N}$ and
$\Psi_{L=0}$ is zero to leading order in $N$: for $\gamma \to
\infty$, the occupation of some ``quasi-momenta" states of
quantum number $\tilde m$ is a flat, Fermi-Dirac-like
distribution, with $|\tilde m| \le N/2$ for $L=0$ \cite{Lieb1}.
To get to the state with $L = N + {\cal O}(1)$, one has to
promote the atom from the single-particle state with $\tilde m
= -N/2$ to the one with $\tilde m = N/2 + 1$. The difference
between the energy of the two states is $(N/2+1)^2 -(-N/2)^2 =
N+1$, which is indeed zero to leading order in $N$ (which is
$N^2$).

For large but finite $\gamma$ the width of the Fermi-Dirac-like
distribution within the Lieb-Liniger model -- shown in the
higher graph of Fig.\,3 -- is $\Delta L = N \gamma /(\gamma +
2) + {\cal O}(1)$ \cite{Lieb1}, or
\begin{eqnarray}
  \Delta L \approx N - 2 \frac N {\gamma} + {\cal O}(1),
\label{deltal}
\end{eqnarray}
neglecting terms of order $N/\gamma^2$ [which is
self-consistent with the final result, that implies that
$\gamma \sim {\cal O}(N)$]. If the single-particle operator
$\Delta V$ acts on the state $|\Psi_{L=N}\rangle$ once, similar
arguments to the ones mentioned in the previous paragraph (for
$\gamma \to \infty$) imply that the change of the total angular
momentum which does not alter the energy of the system to
leading order in $N$, is again equal to the width $\Delta L$.

However, in order for the transition to be allowed to first
order in $\Delta V$, $\Delta L$ has to be larger than the width
of the energy barrier $W$ shown in the lower graph of Fig.\,3
[defined as $W = N - L_b$, where ${\cal E}(L=L_b) = {\cal
E}(L=N)$]. Since for large $\gamma$ the speed of sound for
$L=0$ is $c_s = {\partial {\cal E}}/{\partial L} = N
{\gamma^2}/{(\gamma+2)^2}$ \cite{Lieb2}, therefore ${\cal
E}(L_b) \approx {\cal E}(0) + c_s L_b$, and $L_b = 1 + {\cal O}
(1/\gamma)$, or
\begin{eqnarray}
W = N - 1+ {\cal O}(1/\gamma).
\label{w}
\end{eqnarray}
In order for $W < \Delta L$ (and the decay rate of persistent
currents to be allowed to first order in $\Delta V$),
Eqs.\,(\ref{deltal}) and (\ref{w}) imply that the corresponding
typical value of $\gamma$, denoted as $\gamma_{\rm decay}$, can
be as large as $N$, [or $g_{\rm decay}$ can be as large as
${\cal O} (N^2)$]. In other words, the decay rate of persistent
currents is highly suppressed up to a value of $g_{\rm decay}
\sim {\cal O} (N^2)$, far beyond the coupling $g_{\rm TG} \sim
{\cal O} (N)$ for which the system gets in the Tonks-Girardeau
regime. Over the large window of couplings $g_{\rm TG} \leq g
\leq g_{\rm decay}$ the system is very close to the
Tonks-Girardeau limit and on the same time it supports
persistent currents.

\begin{figure}
\onefigure[width=6.0cm,height=3.7cm]{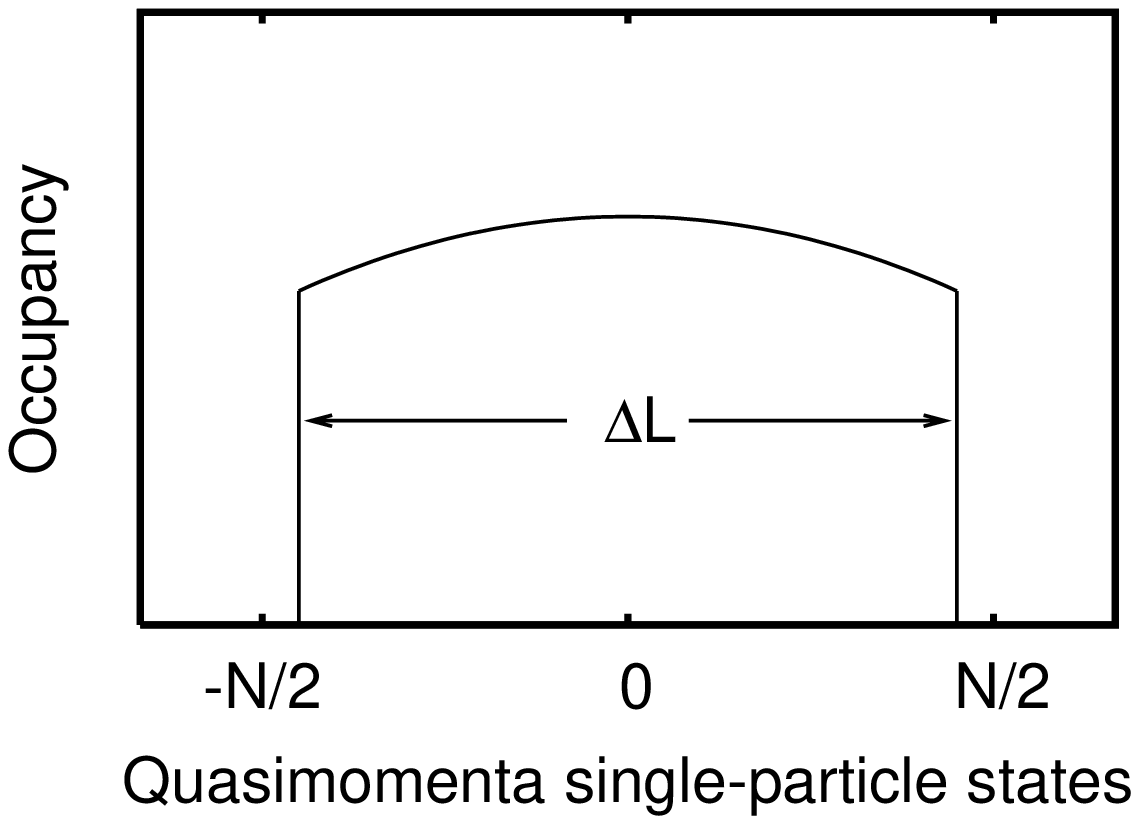}
\onefigure[width=6.0cm,height=3.7cm]{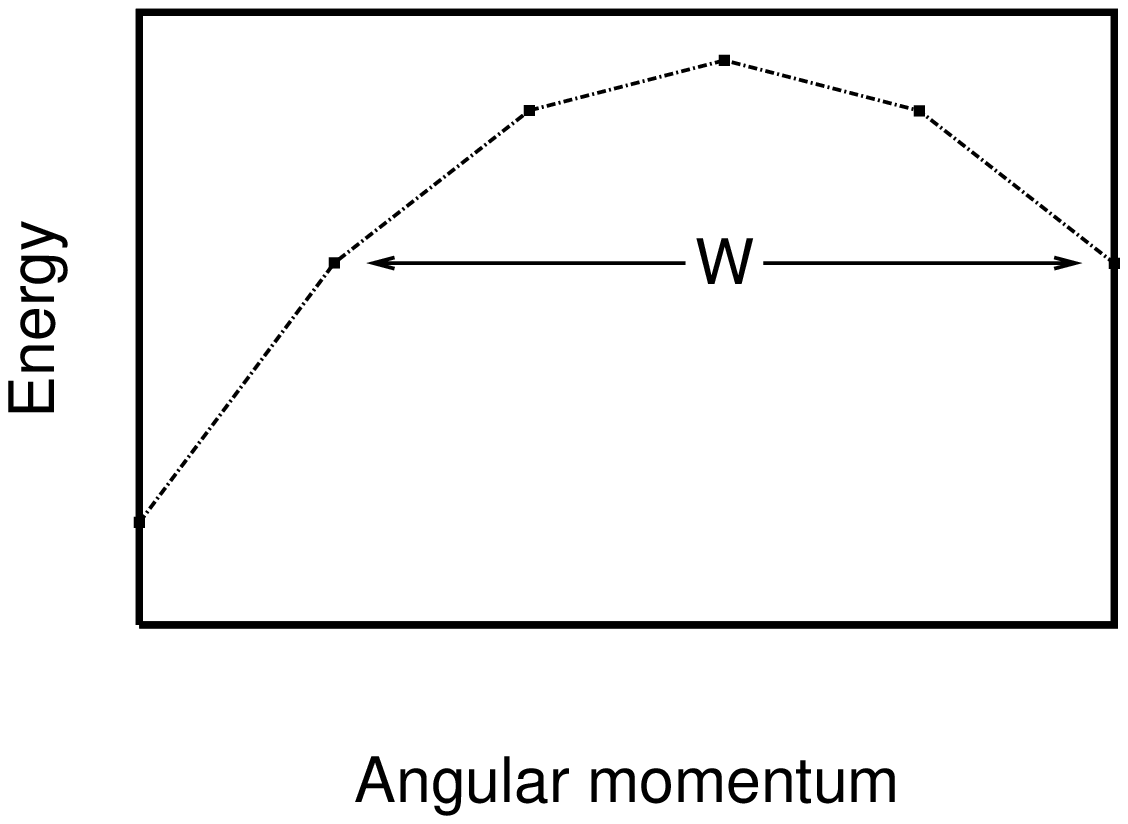}
\caption{Schematic graphs showing $\Delta L$ in the occupancy
of the quasi-momenta states with quantum number $\tilde m$
(high), and $W$ in the dispersion relation ${\cal E} = {\cal
E}(L)$ (low).}
\label{FIG3}
\end{figure}

To get numerical evidence about these estimates, we have
calculated the pair correlation function, defined as
\begin{eqnarray}
   g^{(2)}(\theta - \theta') = \frac {\langle \Psi_{L=0} |
   \sum_{i \neq j} \delta(\theta - \theta_i)
   \delta(\theta' - \theta_j)|
   \Psi_{L=0} \rangle} {(N-1) \langle \Psi_{L=0} | \sum_i
   \delta(\theta - \theta_i) | \Psi_{L=0} \rangle},
\end{eqnarray}
as function of the coupling $g$. Figure 2 shows $1 -
g^{(2)}(0)$ (two higher curves in Fig.\,2) for $N=3$, and 6.
The proximity of $1 - g^{(2)}(0)$ to zero/unity gives the
extent to which the system is in the
Gross-Pitaevskii/Tonks-Girardeau limit. These results are
consistent with the previous estimates $g_{\rm TG} \sim {\cal
O} (N)$, and $g_{\rm decay} \sim {\cal O} (N^2)$. Furthermore,
for a given $g \sim {\cal O} (N)$, the matrix element decreases
with increasing $N$, and probably approaches a step-like
function for very large $N$, that reflects the Fermi-Dirac-like
distribution. Figure 2 provides a specific example in this very
small system, in which for $g_{\rm TG} \leq g \leq g_{\rm
decay}$ the system is close to the Tonks-Girardeau limit, while
the decay rate of persistent currents is suppressed.

\section{Summary} In conclusion, we considered bosonic atoms
which rotate in a toroidal trap and examined the transition
between the Gross-Pitaevskii and the Tonks-Girardeau limits.
For any coupling ${\cal E}(L'=N-L) - {\cal E}(L)=L'-L$, and
thus ${\cal E}(N) - {\cal E}(0) = N$. The energy barrier that
the system has to overcome in order to get from the $l=1$ to
the $l=0$ state and the corresponding distribution function of
the atoms imply that persistent currents exist all the way
between the Gross-Pitaevskii limit -- above some critical
coupling constant -- up to a much higher critical coupling
constant, close to the Tonks-Girardeau limit. Our model
provides an example of a system where Bose-Einstein
condensation is neither a necessary, nor a sufficient condition
for superfluidity.

\acknowledgments
We thank G. Baym, A. Jackson, M. Manninen, and
B. Mottelson for helpful discussions. This work was partly
financed by the European Community project ULTRA-1D
(NMP4-CT-2003-505457), the Swedish Research Council and the
Swedish Foundation for Strategic Research.


\begin{thebibliography} {99}
\bibitem{slope} In the Gross-Pitaevskii limit, when
$L=0$ the gas is in the state $|0^N\rangle$ with an energy $N
(N-1)g/2 + {\cal O}(g)$, while when $L=1$, the gas is in the
state $|0^{N-1}, 1^1\rangle$ with an energy $1+ N (N+1)g/2 +
{\cal O}(g)$. The energy difference is $1 + \delta + {\cal
O}(g)$, in agreement with the slope of $E_{\rm GP}/N$ in
Eq.\,(\ref{Ham8}).
\end{thebibliography}
\end{document}